\documentclass[12pt]{article} 

\usepackage{helvet}

\newcommand*{\myEXPfont}{\fontfamily{cmr}\selectfont}

\usepackage{graphicx} 
\usepackage{mathtools}
\usepackage{amsmath}
\usepackage{multirow}
\usepackage[export]{adjustbox}
\usepackage[left=1in, right=1in, top=1in, bottom=1in]{geometry} 
\usepackage{booktabs}
\usepackage{rotating}
\usepackage{rotfloat}
\usepackage{float}
\usepackage{subfig}
\usepackage{url}
\usepackage{pdfpages}
\usepackage{rotating}
\usepackage[doublespacing]{setspace} 
\usepackage{csvsimple}
\usepackage{array}
\usepackage{wrapfig}
\usepackage[utf8]{inputenc}

\usepackage{cite}
\usepackage{tabularx}
\usepackage{float}
\usepackage{verbatim}

\usepackage{amsthm}                
\usepackage{amssymb}

\usepackage{multirow}

\usepackage{caption}

\newcommand\independent{\protect\mathpalette{\protect\independenT}{\perp}}
\def\independenT#1#2{\mathrel{\rlap{$#1#2$}\mkern2mu{#1#2}}}

\usepackage{amsfonts}
\usepackage{amsmath} 
\usepackage{tikz}
\usetikzlibrary{positioning,shapes.geometric}

\usepackage{xcolor}

\usepackage{setspace}

\DeclareMathOperator{\E}{\mbox{{\myEXPfont E}}}

\usepackage{mathrsfs}

\makeatletter
\newcommand*{\indep}{%
  \mathbin{%
    \mathpalette{\@indep}{}%
  }%
}
\newcommand*{\nindep}{%
  \mathbin{
    \mathpalette{\@indep}{\not}
  }%
}
\newcommand*{\@indep}[2]{%
  \sbox0{$#1\perp\m@th$}
  \sbox2{$#1=$}
  \sbox4{$#1\vcenter{}$}
  \rlap{\copy0}
  \dimen@=\dimexpr\ht2-\ht4-.2pt\relax
  \kern\dimen@
  {#2}%
  \kern\dimen@
  \copy0 
} 
\makeatother

\definecolor{forestgreen}{RGB}{34,139,34}
\usepackage{hyperref}
\hypersetup{
    colorlinks=true,
    linkcolor=magenta,
    filecolor=magenta, 
    citecolor=forestgreen,      
    urlcolor=blue
}

\usepackage{sectsty}
\sectionfont{\large}

\usepackage{layout}

\usepackage{geometry}

\newcolumntype{C}[1]{>{\centering\arraybackslash}p{#1}}

\usepackage{authblk}

\usepackage{afterpage}
\usepackage{lscape}
\usepackage{graphicx}

\usepackage{datetime}
\def\paperversionmajor{6}
\def\paperversionminor{0}

\usepackage{amsthm}

\raggedright
\begin{document}

\title{Generalizing and transporting inferences about the effects of treatment assignment subject to non-adherence}


\author[1-3]{Issa J. Dahabreh\thanks{Address for correspondence: Dr. Issa J. Dahabreh, Department of Epidemiology, Harvard T.H. Chan School of Public Health, Boston, MA 02115; email: \href{mailto:idahabreh@hsph.harvard.edu}{idahabreh@hsph.harvard.edu}; phone: +1 (617) 495‑1000.}}
\author[1,2]{Sarah E. Robertson}
\author[1-3]{Miguel A. Hern\'an}

\affil[1]{CAUSALab, Harvard T.H. Chan School of Public Health, Boston, MA}
\affil[2]{Department of Epidemiology, Harvard T.H. Chan School of Public Health, Boston, MA}
\affil[3]{Department of Biostatistics, Harvard T.H. Chan School of Public Health, Boston, MA}

\maketitle{}
\thispagestyle{empty} 

\thispagestyle{empty}

\clearpage
\vspace*{0.6in}

\thispagestyle{empty}

\begin{abstract}
\noindent
\linespread{1.3}\selectfont
We discuss the identifiability of causal estimands for generalizability and transportability analyses, both under perfect and imperfect adherence to treatment assignment. We consider a setting where the trial data contain information on baseline covariates, assignment at baseline, intervention at baseline (point treatment), and outcomes; and where the data from non-randomized individuals only contain information on baseline covariates. In this setting, we review identification results under perfect adherence and study two examples in which non-adherence severely limits the ability to transport inferences about the effects of treatment assignment to the target population. In the first example, trial participation has a direct effect on treatment receipt and, through treatment receipt, on the outcome (a ``trial engagement effect'' via adherence). In the second example, participation in the trial has unmeasured common causes with treatment receipt. In both examples, the effect of assignment on the outcome in the target population is not identifiable. In the first example, however, the effect of joint interventions to scale-up trial activities that affect adherence and assign treatment is identifiable. We conclude that generalizability and transportability analyses should consider trial engagement effects via adherence and selection for participation on the basis of unmeasured factors that influence adherence.
\end{abstract}

\clearpage
\setcounter{page}{1}
\section*{INTRODUCTION}

Studies extending -- generalizing or transporting \cite{hernan2016discussionkeiding,  dahabreh2019commentaryonweiss} -- inferences about treatment effectiveness from a randomized trial to a target population aim to estimate causal effects in that target population. Recent methodological developments for transportability and generalizability analyses (e.g., \cite{cole2010, westreich2017, rudolph2017, dahabreh2018generalizing, dahabreh2020transportingStatMed}) have focused on addressing differences in the distribution of effect modifiers between the randomized trial population and the target population. Therefore, the proposed methods represent different techniques for standardizing the data distribution of the trial to the covariate distribution of the target population (e.g., using g-formula or weighting approaches \cite{dahabreh2019relation}).

This emphasis on standardization has sidelined important questions about what causal effect (or estimand) is being estimated by analyses extending inferences from the randomized trial to the target population. The most widely reported effect estimates from randomized trials are typically based on comparisons between groups defined by the (randomly) assigned treatment in the trial rather than the treatment actually received. These so-called intention-to-treat effect estimates quantify the effect of treatment assignment \cite{hernan2012beyond} in the trial. Furthermore, to the best of our knowledge, nearly all generalizability and transportability work to-date uses information on the assigned, not the received treatment, in the trial. Thus, transportability and generalizability analyses are arguably aiming to estimate the effects of treatment assignment, that is, intention-to-treat effect analogs, in the target population.

Nevertheless, in the non-experimental context of the target population, the procedures for assignment to treatment may substantially differ from those in the trial, which raises the question of what causal effect is being estimated. Furthermore, the structure and degree of adherence to treatment assignment may greatly differ between the trial and the target population, which may affect the interpretation of the generalizability and transportabiliy analyses \cite{cole2010, westreich2015invited, hong2018generalizing}. In the absence of a precise specification of the causal estimand, we cannot determine whether the standardization procedures that have become so characteristic of transportability analyses are actually estimating any meaningful causal effect.

In this paper, we describe causal estimands that can be of interest in transportability and generalizability analyses. We examine these estimands both under perfect and imperfect adherence to treatment assignment, and discuss the conditions under which the estimands are identifiable. We consider the common setting for such analyses where the trial data contain information on baseline covariates, assignment at baseline, a baseline intervention or point treatment, and outcomes measured at a fixed time of follow-up; and data from non-randomized individuals only contain information on baseline covariates.

\section*{A TRIAL IN A HEALTHCARE SYSTEM} \label{section_setup}

\paragraph{The trial and the target population:} Consider a trial taking place in a healthcare system \cite{olsen2007learning} to study the effectiveness of a baseline intervention (point treatment) on a non-failure time outcome measured at the end of the follow-up. Individuals are invited to participate in the trial; those who agree to participate (and, in many trials, provide informed consent) will be randomly assigned to treatment; those who do not agree, as well as those who meet the trial eligibility criteria but are not invited to participate, will be assigned treatment based on their own and their providers' preferences. 

In the healthcare system, the trial can be embedded in a cohort of individuals for whom the trial is expected to provide information (e.g., individuals for whom the randomized treatments would be used in practice). The cohort can be viewed as a simple random sample from the target population where the trial results will be applied. We have chosen this nested trial design \cite{dahabreh2021studydesigns} for our exposition because it provides the most natural setting for analyses extending causal inferences from the trial to the target population \cite{dahabreh2019commentaryonweiss}; our results, however, also apply to non-nested trial designs, where the trial and the non-randomized subset of the target population are separately sampled \cite{dahabreh2021studydesigns}.  

Nested trial designs can allow, under appropriate conditions, the identification of causal quantities in the target population in which the trial is nested (as opposed to only allowing the identification of causal quantities in subsets of the target population defined by trial participation, as in non-nested trial designs) \cite{dahabreh2021studydesigns}. For example, to address generalizability questions, the target population (and the cohort where the trial is nested) can be restricted to members of the healthcare system who meet all trial eligibility criteria (including, in addition to trial participants, those meeting the trial eligibility criteria but who are not invited to participate, or who are invited but refuse). To address transportability questions, the target population (and the cohort where the trial is nested) can be defined to be broader than the trial eligibility criteria.

\section*{TREATMENT ASSIGNMENT AND TRIAL PARTICIPATION} \label{section_assignment}

Let $X$ denote a vector of baseline covariates, $S$ the trial participation indicator (1 for randomized individuals and 0 for non-randomized individuals), $Z$ the assigned treatment value (which is randomized when $S = 1$), $A$ the received treatment value, and $Y$ the outcome of interest measured at the end of the study. For both randomized and non-randomized individuals in the target population, $Z$ is not equal to $A$ when the individual does not adhere to the assigned treatment.

Because this paper is only concerned with the effect of assigned treatment $Z$, information on received treatment $A$ need not be available for either randomized ($S=1$) or non-randomized ($S=0$) individuals. Furthermore, information on $Z$, $A$, or $Y$ need not be available among non-randomized individuals; in fact, this is the most commonly considered case in generalizability and transportability analyses, where the sample from the non-randomized individuals only contains information on baseline covariates. That is why we refer to the variables in $(S, X, Z, A, Y)$ as observable (i.e., possible to observe): $(S, X)$ are observed in both trial participants and non-participants; $(Z, A, Y)$ are observed for trial participants ($S = 1$) but may be unobserved for non-participants ($S = 0$). To simplify exposition, we assume that there are no losses to follow-up so that the outcome $Y$ is observed on all trial participants.

Our terminology and notation imply that how treatment is assigned does not matter with respect to the outcome of interest. That is, we view the various versions of assignment $Z$ in the randomized trial and in the target population-- say, via an automated randomization system, self-choice, or via consultation with one's medical care provider -- as having the same causal effect on the outcome $Y$.

This assumption of ``variation irrelevance'' \cite{vanderWeele2009} for treatment assignment is implicitly made when intention-to-treat effects estimated in a randomized trial are taken to be informative about the effects of treatment recommendations outside the experimental setting. This assumption, however, is not always reasonable because the assignment process in the experimental context may be fundamentally different from the treatment recommendation process in routine (``real-world'') clinical interactions. This is especially true when treatment assignment involves masking (also referred to as ``blinding'') to prevent trial participants from knowing the treatment they have been assigned to receive, or any other components that cannot occur in the healthcare system.

The assumption of variation irrelevance for treatment assignment may be approximately true for a pragmatic trial. For example, point-of-care randomization may be judged to be sufficiently similar, even if not identical, to modes of prescribing in routine practice (e.g., prescribing via electronic health-record systems). On the other hand, when the assumption does not hold, intention-to-treat effects estimated in the trial may not be adequately informative about the effects of treatment recommendations outside of experimental settings. In that case, other causal estimands (such as the per-protocol effect) may be of greater interest. The results below depend on the assumption that there exits a \emph{hypothetical} intervention to assign treatment in the target population that has the same effect on the outcome as the procedures used to assign treatment in the randomized trial. The practical relevance of learning about such an intervention would depend on the extent to which those assignment procedures can actually be implemented by the healthcare system. 

In addition to variation irrelevance for treatment assignment, we also assume that assignment $Z$ does not affect the outcome $Y$ except through treatment $A$ (e.g., prescription of a medication does not affect the outcome except by influencing treatment). This assumption is made, for example, in trials using the randomized assignment as an instrumental variable \cite{hernan2006instruments}. 

Below, we consider trial engagement effects  \cite{dahabreh2019identification, dahabreh2019commentaryonweiss}, which are effects of trial participation $S$ on the outcome $Y$ that do not operate via the treatment assignment $Z$. Trial engagement effects can occur when there are indirect effects of participation on the outcome via treatment (e.g., by increasing adherence \cite{ horwitz1993adherence, osterberg2005adherence}), or direct effects of participation on the outcome (e.g., Hawthorne effects). Here, we assume that trial engagement effects on the outcome occur only by the impact of trial participation on treatment. Furthermore, we do not distinguish between the invitation to participate and participation itself. Reference \cite{dahabreh2019identification} examines direct engagement effects on the outcome and considers structures that differentiate between trial invitation and participation effects.

\section*{COUNTERFACTUAL MODEL AND IDENTIFIABILITY ASSUMPTIONS} \label{section_counterfactual_model}

To define causal estimands and describe identifiability conditions, we will use counterfactual (potential) outcomes \cite{splawaneyman1990, rubin1974, robins2000d}. Throughout, we adopt a non-parametric structural equation model with finest fully randomized structured tree graph errors \cite{robins1986}. Under this model, counterfactuals are assumed to be well-defined for interventions on any variables and consistency holds for these interventions. Specifically, we will consider the following counterfactual variables to be well-defined: $Z^{s=1}$, the counterfactual assignment under intervention to scale-up trial activities that affect the outcome by setting $S$ to $s=1$; $A^{z}$, the counterfactual treatment under intervention to set the assignment $Z$ to $z$; $A^{s=1, z}$, the counterfactual treatment under intervention to scale-up trial activities and set the assignment $Z$ to $z$; $Y^{z}$, the counterfactual outcome under intervention to set the assignment $Z$ to $z$; and $Y^{s=1, z}$, the counterfactual outcome under intervention to scale-up trial activities to the target population and set the assignment $Z$ to $z$. Under our model the following consistency conditions hold for all individuals in the population under study: if $S = 1$ then $Z^{s=1} = Z$; if $S = 1$ and $Z = z$, then $A^{s=1, z} = A$ and $Y^{s=1, z} = Y$. 

\paragraph{Causal estimands:} An important goal of the study described above is to learn about the effect of assigning the treatments examined in the trial to all members of the target population. That is, we are typically interested in aspects of the distribution of the counterfactual variables, $Y^z$, for the different values of $z$ studied in the trial, and comparisons of the distributions of these counterfactual variables for different $z$ values. For instance, we may be interested in expectations (population means) of counterfactual outcomes in the target population, $\E[Y^z]$, for different values of $z$, as well as average treatment effects comparing assignment to $z$ versus $z^\prime$, defined as $\E[Y^z - Y^{z^\prime}] = \E[Y^z] - \E [ Y^{z^\prime} ]$. 

To the extent that trial engagement effects are present and can be attributed to specific trial activities, we may also be interested in learning about the impact of joint interventions to scale-up these trial activities \emph{and} assign treatment $z$ in the target population. For instance, we may be interested in counterfactual expectations in the target population, $\E[Y^{s=1,z}]$, for different values of $z$, as well as average treatment effects comparing assignment to $z$ versus $z^\prime$, $\E[Y^{s=1, z} - Y^{s=1, z^\prime}] = \E[Y^{s=1, z}] - \E [ Y^{s=1,z^\prime}]$. 

We will now examine the identifiability of these various causal estimands under different causal structures. We have simplified the causal structures in our examples by omitting certain edges which do not affect our results (e.g., we have removed common causes of the unmeasured variables displayed on the graphs and the measured baseline covariates throughout), in order to focus on the most pertinent aspects related to non-adherence.

\paragraph{Positivity conditions:} We assume that the observable variables $(X, S, Z, A)$ have a joint distribution such that (1) for every $x$ with density $f(x) \neq 0$, $1 > \Pr[S = 1 | X = x] > 0$; and (2) for every assigned treatment $z$ evaluated in the trial, and for every $x$ with joint density $f(x,S = 1) \neq 0$, $\Pr[Z = z | X = x, S = 1] > 0$ \cite{dahabreh2020benchmarking, dahabreh2022randomized}.  Of note, our conclusions apply even if the set of \emph{assigned} and \emph{received} treatments, in the trial and outside the trial, are not the same (e.g., if the treatments received by trial participants or non-participants include treatments other than those assigned). 


\section*{PERFECT ADHERENCE TO THE ASSIGNED TREATMENT} 

We first consider the case of perfect adherence, where the received treatment is always the same as the assigned treatment, both in and outside the trial. The causal directed acyclic graph (DAG) of Figure 1A represents the causal structure in the target population (including both trial participants and non-participants). In this section of the paper, we assume that $Z = A$, and denote the deterministic relationship between these two variables using a thicker arrow to connect their corresponding nodes on the DAG (an alternative representation would collapse the two nodes into a single ``treatment'' node).

The baseline covariates affect participation ($X \rightarrow S$), assignment ($X \rightarrow Z$), and, through assignment, treatment ($X \rightarrow Z \rightarrow A$). But once assignment $Z$ is determined, treatment is also determined (because we assume perfect adherence); for that reason $A$ has no other parents except $Z$. The $X \rightarrow Z$ edge indicates that among non-randomized individuals the observed covariates may affect assignment; furthermore, among randomized individuals assignment may be randomized conditional on covariates.

The $S \rightarrow Z$ edge reflects the different mechanisms determining assignment: trial participants are randomized and non-participants are not; the probability of being assigned a specific treatment is usually equal to a constant in the trial (typically, 0.5 when comparing two treatments with 1:1 randomization), whereas it is determined by individual and provider preferences outside the trial. 

The absence of a $S \rightarrow Y$ edge reflects the absence of an effect of participation on the outcome that is not mediated through assignment or treatment receipt (i.e.,  an absence of a direct trial engagement effect). That is because, as noted earlier, we have assumed a causal structure where trial engagement effects are due to the indirect effect of trial participation on the outcome via adherence; we assume that there are no effects of trial participation on the outcome that are not through the assigned treatment. 

The absence of a $Z \rightarrow Y$ edge reflects an assumption that the assignment itself does not affect the outcome except through treatment (e.g., prescription of a medication does not affect the outcome except by influencing treatment). The $U \rightarrow Z$ and $U \rightarrow Y$ edges represent the effects of unmeasured variables that are common causes of assignment, treatment, and the outcome. Of note, common causes of the assignment and the outcome \emph{can} be present in the target population (they operate among non-randomized individuals).

Consider the single world intervention graph (SWIG) \cite{richardson2013single} in Figure 1B representing the intervention to set the treatment assignment $Z$ to $z$. The SWIG obtained from the DAG of Figure 1A represents the intervention on $Z$ by splitting the node (indicated by the vertical line) into a random and fixed part ($Z | z$). The random part receives all incoming edges that were directed into $Z$ in the DAG; the fixed part inherits all outgoing edges emanating from $Z$ in the DAG. All nodes downstream of $z$ are relabeled to represent the counterfactuals under intervention to set $Z$ to $z$ (e.g., $A^z$ and $Y^z$ in Figure 2B). Given the assumption of perfect adherence, if we intervene to set treatment to $Z$ to $z$ then $A^z = z$ for every $z$.

Because of the open path $Z \leftarrow U \rightarrow Y^z$, we do not expect the counterfactual outcomes under assignment $z$ to be independent of the assignment given the baseline covariates, that is, we do not expect the independence $Y^z \indep Z|X$ to hold. Note also that conditioning on $S$ does not eliminate confounding of the effect of assignment on the outcome by $U$: confounding of the effect of assignment on the outcome is expected among non-randomized individuals. 

Some progress is possible by considering the SWIG of Figure 1C under joint interventions of setting $S$ to $s=1$ and $Z$ to $z$ \cite{dahabreh2019identification, dahabreh2020benchmarking}. In this SWIG, we remove the $U \rightarrow Z^{s=1}$ edge because in the world where we have intervened to set $S$ to $s=1$ there are no unmeasured causes of the counterfactual assignment $Z^{s=1}$. We now obtain two independence conditions: $Y^{s=1, z} \indep S|X$ and $Y^{s=1, z} \indep Z^{s=1} | (X, S)$; the second condition, together with consistency, implies the independence condition $Y^{s=1, z} \indep Z | (X, S=1)$. As we show in Web Appendix 1, these conditions, along with positivity and consistency, suffice to identify the expectation of the counterfactual outcome under joint intervention to scale up trial procedures ($s=1$) and assign treatment $z$, $\E[Y^{s=1,z}]$, using a g-formula result \cite{robins1986, dahabreh2019relation, dahabreh2018generalizing}, 
\begin{equation}\label{eq:gformulaID}
    \E[Y^{s=1,z}] = \E\big[\E[Y | X, S = 1 , Z = z]  \big],
\end{equation}
or the equivalent inverse-probability weighting expression, 
\begin{equation}\label{eq:weighitngID} 
    \E\big[\E[Y | X, S = 1 , Z = z]  \big] = \E \left[ \dfrac{I(S = 1, Z = z) Y}{\Pr[S = 1 | X] \Pr[Z = z | X, S = 1]}  \right],
\end{equation}
where $I(S = 1, Z = z)$ is the indicator function that takes value 1, if $S = 1$ and $Z = z$; or 0, otherwise. Thus, under the causal structure of Figure 1A, the counterfactual expectations under joint interventions to scale-up trial activities that influence adherence ($s=1$) and to assign treatment $z$ are identifiable. It follows that the average treatment effects comparing the scaling-up trial activities and assigning treatment $z$ versus $z'$, that is $\E[Y^{s=1, z}-Y^{s=1, z^\prime}]$, are also identifiable. Under the assumption of perfect adherence, we can interpret the effect of assignment as the effect of treatment. 

Under the causal structure of Figure 1A, any effect of trial participation on the outcome is mediated by treatment assignment (there is no direct effect of participation on the outcome); thus, we have that $Y^{s=1,z} = Y^{z}$ for every $z$ and consequently the identification results given above apply to the expectation of the counterfactual outcome under intervention to assign treatment $z$, without reference to scaling-up any trial procedures other than treatment assignment. In other words, under the causal structure of Figure 1A, $\E[Y^{s=1,z}] =  \E[Y^{z}]$ and equations \eqref{eq:gformulaID} and \eqref{eq:weighitngID} hold. 

We now consider two examples that show how imperfect adherence to the assigned treatment severely complicates identification, when trial participation has a direct effect on treatment and when trial participation and adherence have unmeasured common causes.

\section*{NON-ADHERENCE TO THE ASSIGNED TREATMENT} 

\subsection*{Direct effect of participation on treatment} 

Consider now the causal directed acyclic graph (DAG) of Figure 2A for the target population (including both trial participants and non-participants). We no longer assume a deterministic equality between treatment $Z$ and assignment $A$; in fact, non-adherence can be defined as $Z \neq A$. In this example, baseline covariates directly affect participation ($X \rightarrow S$), assignment ($X \rightarrow Z$), treatment ($X \rightarrow A$), and the outcome ($X \rightarrow Y$).

\begin{figure}[htp]
 \caption{DAG and SWIGs with perfect adherence.}
    \centering
    \includegraphics[width=12cm]{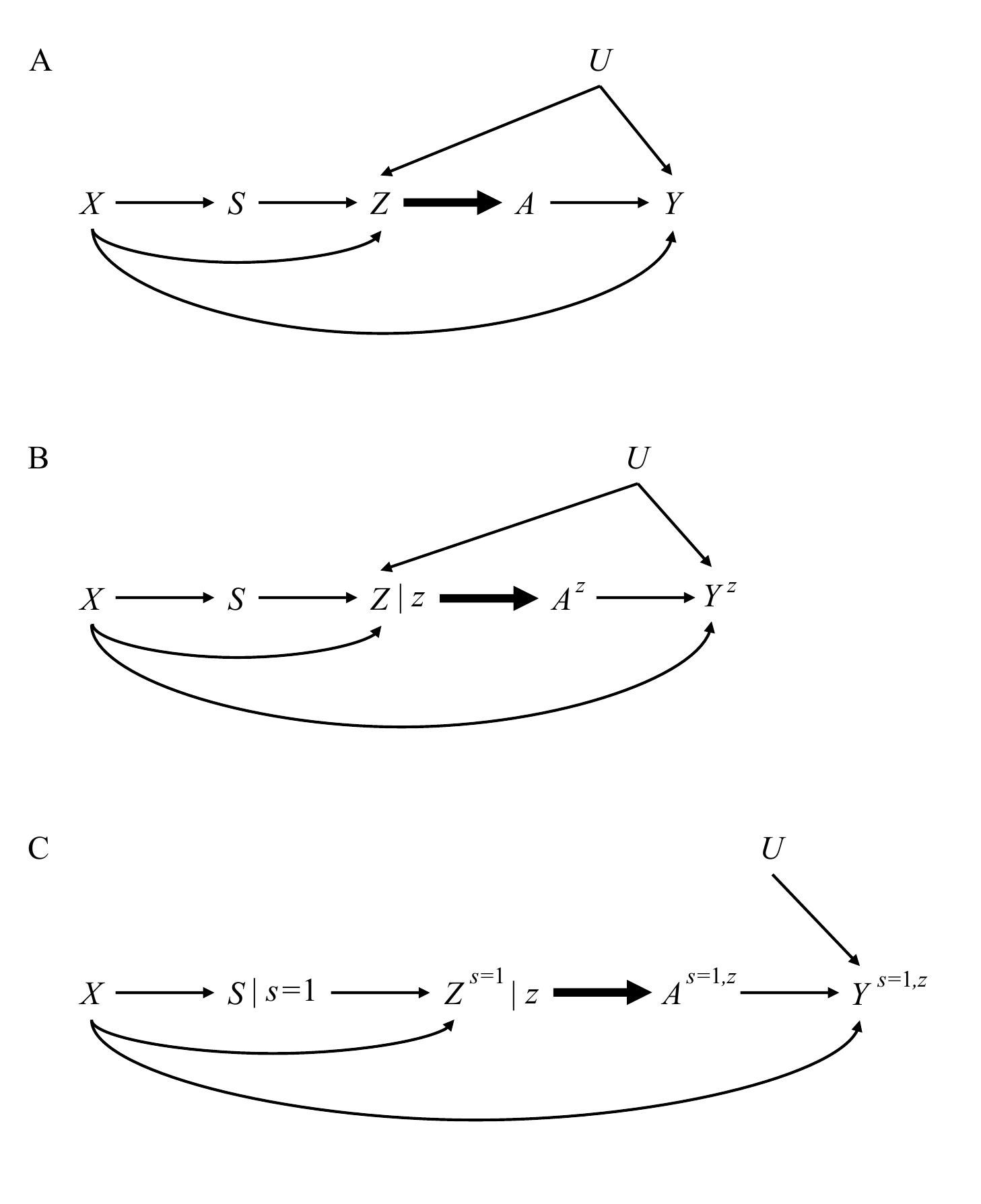}
    \label{fig:graphs_f1}
\end{figure}

As before, the $S \rightarrow Z$ edge reflects the different mechanisms determining assignment: trial participants are randomized and non-participants are not. Under non-adherence, however, the $S \rightarrow A$ edge (which was not present in Figure \ref{fig:graphs_f1}) represents the direct effect of trial participation on the treatment (e.g., when the trial includes routine monitoring visits that encourage adherence to the assigned treatment). When $A$ also affects the outcome, that is, the $A \rightarrow Y$ edge is present, then the $S \rightarrow A \rightarrow Y$ path represents the effect of trial participation on the outcome through treatment (i.e., this path reflects indirect trial engagement effects on the outcome via adherence). As before, the absence of an $S \rightarrow Y$ edge reflects the absence of an effect of participation on the outcome that is not mediated through assignment or treatment receipt (i.e.,  an absence of a direct trial engagement effect). Thus, in the causal structure of Figure 2A, trial engagement effects are due to the indirect effect of trial participation on the outcome via adherence; we assume that there are no effects of trial participation on the outcome that are not through the assigned or received treatment. 

The $Z \rightarrow A$ edge represents the effect of the assignment on treatment (no longer deterministic). The absence of a $Z \rightarrow Y$ edge reflects an assumption that the assignment itself does not affect the outcome except through treatment. The $U \rightarrow Z$, $U \rightarrow A$, and $U \rightarrow Y$ edges represent the effects of unmeasured variables that are common causes of assignment, treatment, and the outcome. Again, common causes of the assignment and the outcome \emph{can} be present among non-randomized individuals in the target population.

Consider the single world intervention graph (SWIG) \cite{richardson2013single} in Figure 2B representing the intervention to set the treatment assignment $Z$ to $z$, obtained from the DAG of Figure 2A. Because of the open paths $Z \leftarrow U \rightarrow Y^z$, $Z \leftarrow U \rightarrow A^z \rightarrow Y^z$, and $Z \leftarrow S \rightarrow A^z \rightarrow Y^z$, we do not expect the counterfactual outcomes under assignment $z$ to be independent of the assignment given the baseline covariates, that is, we do not expect the independence $Y^z \indep Z|X$ to hold. As in the perfect adherence case, conditioning on $S$ does not eliminate confounding of the effect of assignment on the outcome by $U$. 

Similar to the case under perfect adherence, progress is possible by considering the SWIG of Figure 2C under joint interventions of setting $S$ to $s=1$ and $Z$ to $z$ \cite{dahabreh2019identification, dahabreh2020benchmarking}. In this SWIG, we remove the $U \rightarrow Z^{s=1}$ edge because in the world where we have intervened to set $S$ to $s=1$ there are no unmeasured causes of the corresponding counterfactual assignment $Z^{s=1}$. We now obtain the same two independence conditions as under perfect adherence: $Y^{s=1, z} \indep S|X$ and $Y^{s=1, z} \indep Z^{s=1} | (X, S)$; the second condition, together with consistency, implies the independence condition $Y^{s=1, z} \indep Z | (X, S=1)$. These conditions, along with positivity and consistency, again suffice to identify $\E[Y^{s=1,z}]$, using the same g-formula and inverse-probability weighting expression given in equations \eqref{eq:gformulaID} and \eqref{eq:weighitngID}, respectively \cite{robins1986, dahabreh2019relation, dahabreh2018generalizing}. The proof of these identification results is the same as in the case of perfect adherence. Thus, under the causal structure of Figure 2A, the counterfactual expectations under joint interventions to scale-up trial activities that influence adherence ($s=1$) and to assign treatment $z$ are identifiable, even when adherence is imperfect. It follows that the average treatment effects comparing to scale-up trial activities and assign treatment $z$ versus $z'$, that is $\E[Y^{s=1, z}-Y^{s=1, z}]$, are also identifiable. 

Despite the similarities between our analysis of the causal structures of Figure 1 and Figure 2 up to this point, a major difference arises in the presence of non-adherence: the $S \rightarrow A \rightarrow Y$ path on the DAG means that in general we do not expect $Y^z$ to equal $Y^{s=1, z}$. Thus, when trial participation affects the outcome through treatment, we do not expect the counterfactual expectation under intervention to assign treatment $z$, that is, $\E[Y^{z}]$, to equal the counterfactual expectation to scale-up trial activities that influence adherence and assign treatment, $\E[Y^{s=1,z}]$. Additionally, we do not expect the average treatment effects comparing assignment to treatment $z$ and $z^\prime$, $\E[Y^{z} - Y^{z^\prime}]$, to equal the average treatment effects to scale-up trial activities that influence adherence and assign treatment, $\E[Y^{s=1, z} - Y^{s=1, z^\prime}]$. In fact, $\E[Y^{z}]$ and $\E[Y^{z} - Y^{z^\prime}]$ are not identifiable in this causal structure: the effect of assignment is confounded among non-randomized individuals; and randomized individuals only provide information about the effect of both scaling-up trial activities and assigning treatment because trial participation indirectly affects the outcome (by its direct effect on treatment). Informally, the issue is that trial participation affects the outcome through adherence, thus when using outcome information from trial participants we cannot separate the effect of trial participation from that of assignment (without additional assumptions). Of note, this issue would persist even if treatment and the outcome were unconfounded (i.e., even in a causal structure resulting from removing the $A \leftarrow U \rightarrow Y$ fork from the DAG of Figure 2A). 

Consequently, under the DAG of Figure 2A (and the positivity and consistency conditions), the g-formula or weighting approaches from the previous section identify the impact of scaling-up trial activities and setting the treatment assignment, such as $\E[Y^{s=1,z}]$, but not the impact of assignment alone, such as $\E[Y^{z}]$. 

The fact that $\E[Y^{z}]$ and $\E[Y^{z} - Y^{z^\prime}]$ are unidentifiable, while $\E[Y^{s=1,z}]$ and $\E[Y^{s=1, z} - Y^{s=1, z^\prime}]$ are identifiable under the DAG of Figure 2A, is probably what Hong et al. meant when they stated that their estimate was ``...based on the trial population, including any effect of the trial on adherence and persistence...'' \cite{hong2018generalizing}. As we will see next, however, both kinds of estimands may be unidentifiable when trial participation has unmeasured common causes with the receipt of treatment.

\subsection*{Common causes of participation and treatment} 

Consider next the DAG of Figure 3A, obtained from the DAG of Figure 2A by removing the $S \rightarrow A$ edge, adding a common cause of participation and treatment (the $ S \leftarrow U_1 \rightarrow A$ fork), and substituting $U_2$ for $U$ (to keep the notation consistent); the rest of the structure is the same between the two DAGs. In this DAG, $U_1$ represents unobserved common causes of trial participation and treatment. For example, participation in the trial may share common causes with treatment when trial participants tend to be more health-conscious and willing to adhere to the assigned treatment. Information on characteristics such as health-consciousness is typically not collected in randomized trials and is unavailable in most target population samples.

In the SWIG of Figure 3B under intervention to set $Z$ to $z$, we do not expect the independence $Y^z \indep Z|X$ to hold because of the open paths $Z \leftarrow S \leftarrow U_1 \rightarrow A^z \rightarrow Y^z$, $Z \leftarrow U_2 \rightarrow Y^z$, and $Z \leftarrow U_2 \rightarrow A^z \rightarrow Y^z$. As before, to make progress we consider joint interventions to scale-up trial activities related to adherence and assign treatment.

In the SWIG of Figure 3C under joint intervention to set $S$ to $s=1$ and $Z$ to $z$, we remove the $U_2 \rightarrow Z^{s=1}$ edge because in the world where we have intervened to set $S$ to $s=1$ there are no unmeasured causes of the corresponding counterfactual assignment $Z^{s=1}$, similar to the SWIGs of Figure 1C and 2C. We again obtain the independence $Y^{s=1,z} \indep Z^{s=1}|(X,S)$, which, along with consistency, implies $Y^{s=1, z} \indep Z | (X, S=1)$. In contrast to the SWIG of Figures 1C and 2C, however, in the SWIG of Figure 3C we do not expect the independence $Y^{s=1,z} \indep S|X$ to hold because of the open $S \leftarrow U_1 \rightarrow A^{s=1,z} \rightarrow Y^{s=1,z}$ path. Note also that conditioning on $A^{s=1,z}$ cannot help because $A^{s=1,z}$ is a collider on the $S \leftarrow U_1 \rightarrow A^{s=1,z} \leftarrow U_2 \rightarrow Y^{s=1,z}$ path. Consequently, we cannot identify counterfactual expectations under joint interventions to scale-up trial activities and assign treatment $z$, $\E[Y^{s=1,z}]$ because the effect of assignment is confounded among non-randomized individuals; and randomized individuals are not exchangeable with non-randomized individuals (with respect to the counterfactual outcomes outcome $Y^{s=1,z}$) because trial participation is driven by unmeasured variables that also affect the outcome through treatment. Furthermore, we cannot identify average treatment effects comparing assignments $z$ and $z'$, that is, $\E[Y^{s=1,z}-Y^{s=1,z}]$. Informally, the issue is that trial participants are selected for unmeasured variables that influence (usually, increase) adherence compared with non-participants and adherence affects the outcome; thus, the information from trial participants mixes the effect of treatment assignments with the effect of the unmeasured variables.

Under the causal structure of Figure 3A, $Y^z$ is equal to $Y^{s=1, z}$ because the $S \rightarrow A \rightarrow Y$ path has been removed (this can be seen by redrawing the SWIG of Figure 3C using minimal labeling \cite{richardson2013single}). Thus, we expect that $\E[Y^{z}]=\E[Y^{s=1,z}]$ and $\E[Y^{z} - Y^{z^\prime}]=\E[Y^{s=1, z} - Y^{s=1, z^\prime}]$. This, however, is not helpful because, as we discussed, all these quantities are unidentifiable. 

Finally, consider a DAG that combines the edges of the DAGs in Figures 2A and 3A, such that both a common cause of trial participation and a direct effect of participation on the received treatment are present (see Web Appendix Figure 2). Clearly, using the same arguments as above, $\E[Y^{z}]$ and $\E[Y^{s=1,z}]$, as well as average treatment effects based on contrasts of these expectations for different $z$ values, are unidentifiable under this causal structure.

\section*{OTHER CAUSAL ESTIMANDS} 
For completeness, we now briefly consider estimands other than those that pertain to the effect of assignment on the outcome in the target population.

\paragraph{Effects in the non-randomized subset of the target population:} Estimands that pertain to the \textit{non-randomized subset} of the target population may be of interest in nested trial designs, in addition to those that pertain to the entire target population  \cite{dahabreh2020transportingStatMed}. These estimands are important because they are also identifiable (provided certain sampling conditions hold) in non-nested trial designs where data from a trial are combined with data from a separately obtained sample of non-randomized individuals from the target population \cite{dahabreh2021studydesigns} (in non-nested trial designs, estimands in the entire target population are not identifiable without additional information). In Web Appendix 2 we show that the findings from our examples apply with minimal modifications to estimands for the non-randomized subset of the target population, both for nested or non-nested trial designs.

\paragraph{Per-protocol effects:} In all the examples in this paper, per-protocol effects in the target population (i.e., effects of treatment according to assignment) are unidentifiable because of intractable confounding of the effect of treatment $A$ on the outcome $Y$, indicated by the $A \leftarrow U_2 \rightarrow Y$ fork in Figure 3A ($U$ in Figure 2A). For the same reason, the effects of joint interventions to scale-up trial activities that influence adherence and set the treatment to $a$ are also unidentifiable. 

\paragraph{The effect of assignment in the population underlying the trial:} As is well known \cite{hernan2020}, the counterfactual means of assignment $z$ in the population underlying the trial (i.e., the randomized subset of the target population), $\E[Y^{z}|S=1]$, as well as average treatment effects comparing assignment to treatment $z$ versus $z^\prime$, $\E[Y^{z} - Y^{z^\prime}|S=1]$, are identifiable under all the DAGs we have considered in this paper. For example, using the DAG of Appendix Figure 1A (see Web Appendix 3), which combines the complications we examined in the DAGs of Figures 2A and 3A, and is conditional on $S=1$, we can construct the SWIG of Appendix Figure 1B, under intervention to set $Z$ to $z$. Because we condition on $S=1$, the $U_2 \rightarrow Z$ edge can be removed (in the trial, the assignment $Z$ does not depend on unmeasured variables) and we obtain the independence condition $ Y^z \independent Z | (X, S = 1)$. As we show in Web Appendix 3, this independence condition, and the positivity and consistency conditions, suffice to identify the effect of treatment assignment in the population underlying the trial: $ \E[Y^z | S = 1] = \E \big[ \E[Y | X, S = 1, Z = z] \big | S = 1 \big]$.

\section*{DISCUSSION}

Extending causal inferences from one population to another requires similarities between them that extend beyond having the same distribution of baseline effect modifiers (or outcome predictors) \cite{hernan2011compound}. Yet, most work on generalizing or transporting causal inferences has focused exclusively on adjusting for differences in the distributions of baseline covariates between trial participants and non-participants. Here, we considered estimands for extending inferences to a target population when the trial data provide information on baseline covariates, the randomized treatment assignments, and outcomes; and when the data from the target population only provide information on baseline covariates. By means of studying stylized examples, we showed that the interpretation of generalizability and transportability analyses is complicated when trial engagement affects the outcome via adherence and when participation in the trial and treatment receipt share common causes.

Our examples suggest that generalizability and transportability analyses that standardize the data from a trial to the covariate distribution of the target population using information only on assigned treatment in the trial, without information on adherence, may produce estimates that do not have a causal interpretation as effects of treatment assignment in the target population when trial participation and treatment have unmeasured common causes. In the presence of such common causes, the effect of assignment is not identifiable. When there are no unmeasured common causes of trial participation and treatment, however, the results of conventional standardization methods can be given a new interpretation as estimating the effects of joint interventions to scale-up trial activities that influence adherence and to assign treatment. 

Our analysis highlights that the causal interpretation of generalizability and transportability analyses that standardize the trial data to the target population depends on characteristics of the populations and aspects of trial design. For instance, endowing these approaches with a causal interpretation may not be appropriate for confirmatory phase III clinical trials, which typically involve activities to promote adherence and select participants who are likely to be adherent, or when substantial non-adherence is expected. In contrast, our findings are less troubling for highly pragmatic trials (and observational analyses emulating pragmatic target trials \cite{hernan2016}) and in settings where near-perfect adherence is expected. 

Throughout, we focused on identification under various distributional independence (exchangeability) conditions between the counterfactual outcomes and trial participation or assignment because these conditions can be directly read off SWIGs. Identification is sometimes possible under weaker conditions of conditional exchangeability in expectation or in effect measure rather than in distribution (e.g., \cite{dahabreh2018generalizing,  dahabreh2020transportingStatMed}). Despite being weaker than the independence conditions encoded in our SWIGs, the plausibility of independence in conditional expectation or effect measure is still severely reduced when trial participation affects the outcome through treatment (the $S \rightarrow A \rightarrow Y$ path of Figure 2A) or common causes of trial participation and the outcome (the $S \leftarrow U_1 \rightarrow A \rightarrow Y$ path of Figure 3A).

For simplicity, we limited our attention to point treatments and did not consider sustained treatments. Nevertheless, the problems we identified are likely more pressing in generalizability and transportability analyses of trials comparing sustained treatments. Patterns of adherence for sustained treatments are more complex and adherence rates are often very low (e.g., up to half of individuals prescribed statins quit within a year \cite{brown2011medication}). In trials of sustained treatments, trial engagement effects on the outcome via adherence may be more likely when the trials involve regular protocol-directed contact between study participants and personnel, which may increase adherence \cite{petrilla2005evidence}. Furthermore, individuals that participate in long-term trials of sustained treatments may be unrepresentative of those seen in general practice in terms of their propensity to adhere to treatment.

Our examples were highly stylized to focus on core issues related to non-adherence. A more refined representation of the trial invitation and participation process \cite{dahabreh2019identification, dahabreh2019commentaryonweiss} would, if anything, lead to additional possibilities for trial engagement effects or the existence of unmeasured common causes between invitation or participation, and treatment. Furthermore, we treated trial participation as a well-defined intervention that would include all aspects of the trial other than random treatment assignment \cite{dahabreh2019identification}. In practice, understanding the relationship between trial participation and adherence will require more refined causal models and data collection about specific trial activities that affect patient behaviors. 

Despite these limitations, our analysis highlights that framing problems of generalizability and transportability of treatment effects as simple problems of standardization to the baseline covariate distribution of the target population can obscure the choice of causal estimands and the assumptions needed to identify them. Instead, generalizability and transportability problems should be treated as proper causal problems, in the sense that estimand choice and identifiability analysis are best served by explicitly considering the underlying causal structure. In our simple examples, the structural approach highlighted the importance of trial engagement effects on the outcome via adherence and selective participation into the trial on the basis of unmeasured variables that also influence adherence to the assigned treatment.

\section*{ACKNOWLEDGMENTS}

This section has been temporarily removed from the manuscript for peer review.

\clearpage
\renewcommand{\refname}{REFERENCES}
\bibliographystyle{ieeetr}
\bibliography{bibliography}

\begin{thebibliography}{10}

\bibitem{hernan2016discussionkeiding}
M.~A. Hern{\'a}n, ``Discussion of ``{P}erils and potentials of self-selected
  entry to epidemiological studies and surveys'','' {\em Journal of the Royal
  Statistical Society. Series A (Statistics in Society)}, vol.~179, no.~2,
  pp.~346--347, 2016.

\bibitem{dahabreh2019commentaryonweiss}
I.~J. Dahabreh and M.~A. Hern{\'a}n, ``Extending inferences from a randomized
  trial to a target population,'' {\em European Journal of Epidemiology},
  vol.~34, no.~8, pp.~719--722, 2019.

\bibitem{cole2010}
S.~R. Cole and E.~A. Stuart, ``Generalizing evidence from randomized clinical
  trials to target populations: the {A}{C}{T}{G} 320 trial,'' {\em American
  {J}ournal of {E}pidemiology}, vol.~172, no.~1, pp.~107--115, 2010.

\bibitem{westreich2017}
D.~Westreich, J.~K. Edwards, C.~R. Lesko, E.~Stuart, and S.~R. Cole,
  ``Transportability of trial results using inverse odds of sampling weights,''
  {\em American Journal of Epidemiology}, vol.~186, no.~8, pp.~1010--1014,
  2017.

\bibitem{rudolph2017}
K.~E. Rudolph and M.~J. van~der Laan, ``Robust estimation of encouragement
  design intervention effects transported across sites,'' {\em Journal of the
  Royal Statistical Society. Series B (Statistical Methodology)}, vol.~79,
  no.~5, pp.~1509--1525, 2017.

\bibitem{dahabreh2018generalizing}
I.~J. Dahabreh, S.~E. Robertson, E.~J. Tchetgen~Tchetgen, E.~A. Stuart, and
  M.~A. Hern{\'a}n, ``Generalizing causal inferences from individuals in
  randomized trials to all trial-eligible individuals,'' {\em Biometrics},
  vol.~75, no.~2, pp.~685--694, 2018.

\bibitem{dahabreh2020transportingStatMed}
I.~J. Dahabreh, S.~E. Robertson, J.~A. Steingrimsson, E.~A. Stuart, and M.~A.
  Hern{\'a}n, ``Extending inferences from a randomized trial to a new target
  population,'' {\em Statistics in Medicine}, vol.~39, no.~14, pp.~1999--2014,
  2020.

\bibitem{dahabreh2019relation}
I.~J. Dahabreh, S.~E. Robertson, and M.~A. Hern{\'a}n, ``On the relation
  between g-formula and inverse probability weighting estimators for
  generalizing trial results,'' {\em Epidemiology}, vol.~30, no.~6,
  pp.~807--812, 2019.

\bibitem{hernan2012beyond}
M.~A. Hern{\'a}n and S.~Hern{\'a}ndez-D{\'\i}az, ``Beyond the
  intention-to-treat in comparative effectiveness research,'' {\em Clinical
  Trials}, vol.~9, no.~1, pp.~48--55, 2012.

\bibitem{westreich2015invited}
D.~Westreich and J.~K. Edwards, ``Invited commentary: every good randomization
  deserves observation,'' {\em American Journal of Epidemiology}, vol.~182,
  no.~10, pp.~857--860, 2015.

\bibitem{hong2018generalizing}
J.-L. Hong, M.~Jonsson~Funk, R.~LoCasale, S.~E. Dempster, S.~R. Cole,
  M.~Webster-Clark, J.~K. Edwards, and T.~St{\"u}rmer, ``Generalizing
  randomized clinical trial results: implementation and challenges related to
  missing data in the target population,'' {\em American Journal of
  Epidemiology}, vol.~187, no.~4, pp.~817--827, 2018.

\bibitem{olsen2007learning}
``The learning healthcare system: workshop summary,'' in {\em {IOM} roundtable
  on evidence-based medicine} (L.~Olsen, D.~Aisner, and J.~M. McGinnis, eds.),
  DC: National Academies Press, 2007.

\bibitem{dahabreh2021studydesigns}
I.~J. Dahabreh, S.~J.-P. Haneuse, J.~M. Robins, S.~E. Robertson, A.~L.
  Buchanan, E.~A. Stuart, and M.~A. Hern\'an, ``Study designs for extending
  causal inferences from a randomized trial to a target population,'' {\em
  American Journal of Epidemiology}, vol.~190, no.~8, pp.~1632--1642, 2021.

\bibitem{vanderWeele2009}
T.~J. VanderWeele, ``Concerning the consistency assumption in causal
  inference,'' {\em Epidemiology}, vol.~20, no.~6, pp.~880--883, 2009.

\bibitem{hernan2006instruments}
M.~A. Hern{\'a}n and J.~M. Robins, ``Instruments for causal inference: an
  epidemiologist's dream?,'' {\em Epidemiology}, pp.~360--372, 2006.

\bibitem{dahabreh2019identification}
I.~J. Dahabreh, J.~M. Robins, S.~J.-P. Haneuse, and M.~A. Hern\'an,
  ``Generalizing causal inferences from randomized trials: counterfactual and
  graphical identification,'' {\em arXiv preprint arXiv:1906.10792}, 2019
  (accessed: 11/03/2020).

\bibitem{horwitz1993adherence}
R.~I. Horwitz and S.~M. Horwitz, ``Adherence to treatment and health
  outcomes,'' {\em Archives of Internal Medicine}, vol.~153, no.~16,
  pp.~1863--1868, 1993.

\bibitem{osterberg2005adherence}
L.~Osterberg and T.~Blaschke, ``Adherence to medication,'' {\em New England
  Journal of Medicine}, vol.~353, no.~5, pp.~487--497, 2005.

\bibitem{splawaneyman1990}
J.~Splawa-Neyman, ``On the application of probability theory to agricultural
  experiments. essay on principles. section 9. [{T}ranslated from
  {S}plawa-{N}eyman, {J} (1923) in {R}oczniki {N}auk {R}olniczych {T}om {X},
  1--51],'' {\em Statistical Science}, vol.~5, no.~4, pp.~465--472, 1990.

\bibitem{rubin1974}
D.~B. Rubin, ``Estimating causal effects of treatments in randomized and
  nonrandomized studies.,'' {\em Journal of {E}ducational {P}sychology},
  vol.~66, no.~5, p.~688, 1974.

\bibitem{robins2000d}
J.~M. Robins and S.~Greenland, ``Causal inference without counterfactuals:
  comment,'' {\em Journal of the American Statistical Association}, vol.~95,
  no.~450, pp.~431--435, 2000.

\bibitem{robins1986}
J.~M. Robins, ``A new approach to causal inference in mortality studies with a
  sustained exposure period -- application to control of the healthy worker
  survivor effect,'' {\em Mathematical Modelling}, vol.~7, no.~9,
  pp.~1393--1512, 1986.

\bibitem{dahabreh2020benchmarking}
I.~J. Dahabreh, J.~M. Robins, and M.~A. Hern{\'a}n, ``Benchmarking
  observational methods by comparing randomized trials and their emulations,''
  {\em Epidemiology}, vol.~31, no.~5, pp.~614--619, 2020.

\bibitem{dahabreh2022randomized}
I.~J. Dahabreh, J.~A. Steingrimsson, J.~M. Robins, and M.~A. Hern{\'a}n,
  ``Randomized trials and their observational emulations: a framework for
  benchmarking and joint analysis,'' {\em arXiv preprint arXiv:2203.14857},
  2022.

\bibitem{richardson2013single}
T.~S. Richardson and J.~M. Robins, ``Single world intervention graphs
  ({S}{W}{I}{G}s): A unification of the counterfactual and graphical approaches
  to causality,'' Tech. Rep. 128, Center for Statistics and the Social
  Sciences, University of Washington, 2013 (accessed: 11/03/2020).
\newblock
  \url{https://www.csss.washington.edu/research/working-papers/single-world-intervention-graphs-swigs-unification-counterfactual-and}.

\bibitem{hernan2020}
M.~A. Hern{\'a}n and J.~M. Robins, {\em Causal inference (forthcoming)}.
\newblock Boca Raton, FL: Chapman \& {H}all/{C}{R}{C}, 2020.

\bibitem{hernan2011compound}
M.~A. Hern{\'a}n and T.~J. VanderWeele, ``Compound treatments and
  transportability of causal inference,'' {\em Epidemiology (Cambridge,
  Mass.)}, vol.~22, no.~3, p.~368, 2011.

\bibitem{hernan2016}
M.~A. Hern{\'a}n and J.~M. Robins, ``Using {B}ig {D}ata to emulate a target
  trial when a randomized trial is not available,'' {\em American {J}ournal of
  {E}pidemiology}, vol.~183, no.~8, pp.~758--764, 2016.

\bibitem{brown2011medication}
M.~T. Brown and J.~K. Bussell, ``Medication adherence: {WHO} cares?,'' {\em
  Mayo Clinic Proceedings}, vol.~86, no.~4, pp.~304--314, 2011.

\bibitem{petrilla2005evidence}
A.~Petrilla, J.~Benner, D.~Battleman, J.~Tierce, and E.~Hazard,
  ``Evidence-based interventions to improve patient compliance with
  antihypertensive and lipid-lowering medications,'' {\em International Journal
  of Clinical Practice}, vol.~59, no.~12, pp.~1441--1451, 2005.

\end{thebibliography}


\begin{thebibliography}{1}

\bibitem{dahabreh2019relation}
Issa~J Dahabreh, Sarah~E Robertson, and Miguel~A Hern{\'a}n.
\newblock On the relation between g-formula and inverse probability weighting
  estimators for generalizing trial results.
\newblock {\em Epidemiology}, 30(6):807--812, 2019.

\bibitem{dahabreh2019identification}
Issa~J Dahabreh, James~M Robins, Sebastien J-PA Haneuse, and Miguel~A Hern\'an.
\newblock Generalizing causal inferences from randomized trials: counterfactual
  and graphical identification.
\newblock {\em arXiv preprint arXiv:1906.10792}, 2019 (accessed: 11/03/2020).

\bibitem{richardson2013single}
Thomas~S Richardson and James~M Robins.
\newblock Single world intervention graphs ({S}{W}{I}{G}s): A unification of
  the counterfactual and graphical approaches to causality.
\newblock Technical Report 128, Center for Statistics and the Social Sciences,
  University of Washington, 2013 (accessed: 11/03/2020).
\newblock
  \url{https://www.csss.washington.edu/research/working-papers/single-world-intervention-graphs-swigs-unification-counterfactual-and}.

\bibitem{dahabreh2020transportingStatMed}
Issa~J Dahabreh, Sarah~E Robertson, Jon~A Steingrimsson, Elizabeth~A Stuart,
  and Miguel~A Hern{\'a}n.
\newblock Extending inferences from a randomized trial to a new target
  population.
\newblock {\em Statistics in Medicine}, 39(14):1999--2014, 2020.

\bibitem{dahabreh2021studydesigns}
Issa~J Dahabreh, Sebastien J-PA Haneuse, James~M Robins, Sarah~E Robertson,
  Ashley~L Buchanan, Elisabeth~A Stuart, and Miguel~A Hern\'an.
\newblock Study designs for extending causal inferences from a randomized trial
  to a target population.
\newblock {\em American Journal of Epidemiology}, 190(8):1632--1642, 2021.

\bibitem{bickel1993efficient}
Peter~J Bickel, Chris~AJ Klaassen, Jon~A Wellner, and Ya'acov Ritov.
\newblock {\em Efficient and adaptive estimation for semiparametric models}.
\newblock Johns Hopkins University Press Baltimore, 1993.

\bibitem{hernan2006estimating}
Miguel~A Hern{\'a}n and James~M Robins.
\newblock Estimating causal effects from epidemiological data.
\newblock {\em Journal of Epidemiology \& Community Health}, 60(7):578--586,
  2006.

\end{thebibliography}


\ddmmyyyydate 
\newtimeformat{24h60m60s}{\twodigit{\THEHOUR}.\twodigit{\THEMINUTE}.32}
\settimeformat{24h60m60s}
\begin{center}
\vspace{\fill}\ \newline
\textcolor{black}{{\tiny $ $generalizability\_with\_nonadherence, $ $ }
{\tiny $ $Date: \today~~ \currenttime $ $ }
{\tiny $ $Revision: \paperversionmajor.\paperversionminor $ $ }}
\end{center}

\clearpage
\section*{Figures}

\begin{figure}[htp]
 \caption{DAG and SWIGs for direct effect of participation on adherence.}
    \centering
   \includegraphics[width=12cm]{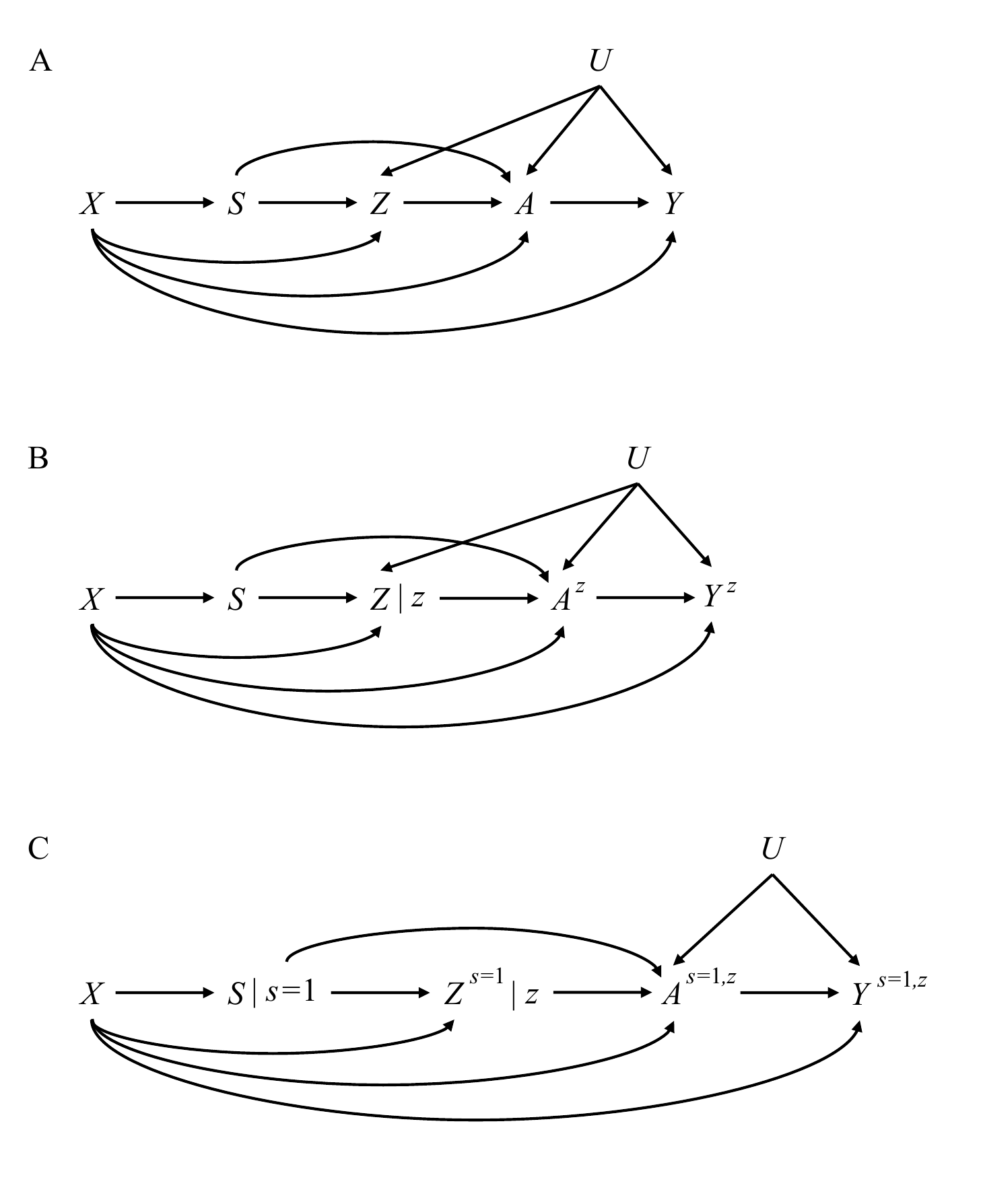}
    \caption*{}
    \label{fig:graphs_f2}
\end{figure}

\clearpage

\begin{figure}[htp]
 \caption{DAG and SWIGs for common causes of trial participation and treatment.}
    \centering
   \includegraphics[width=12cm]{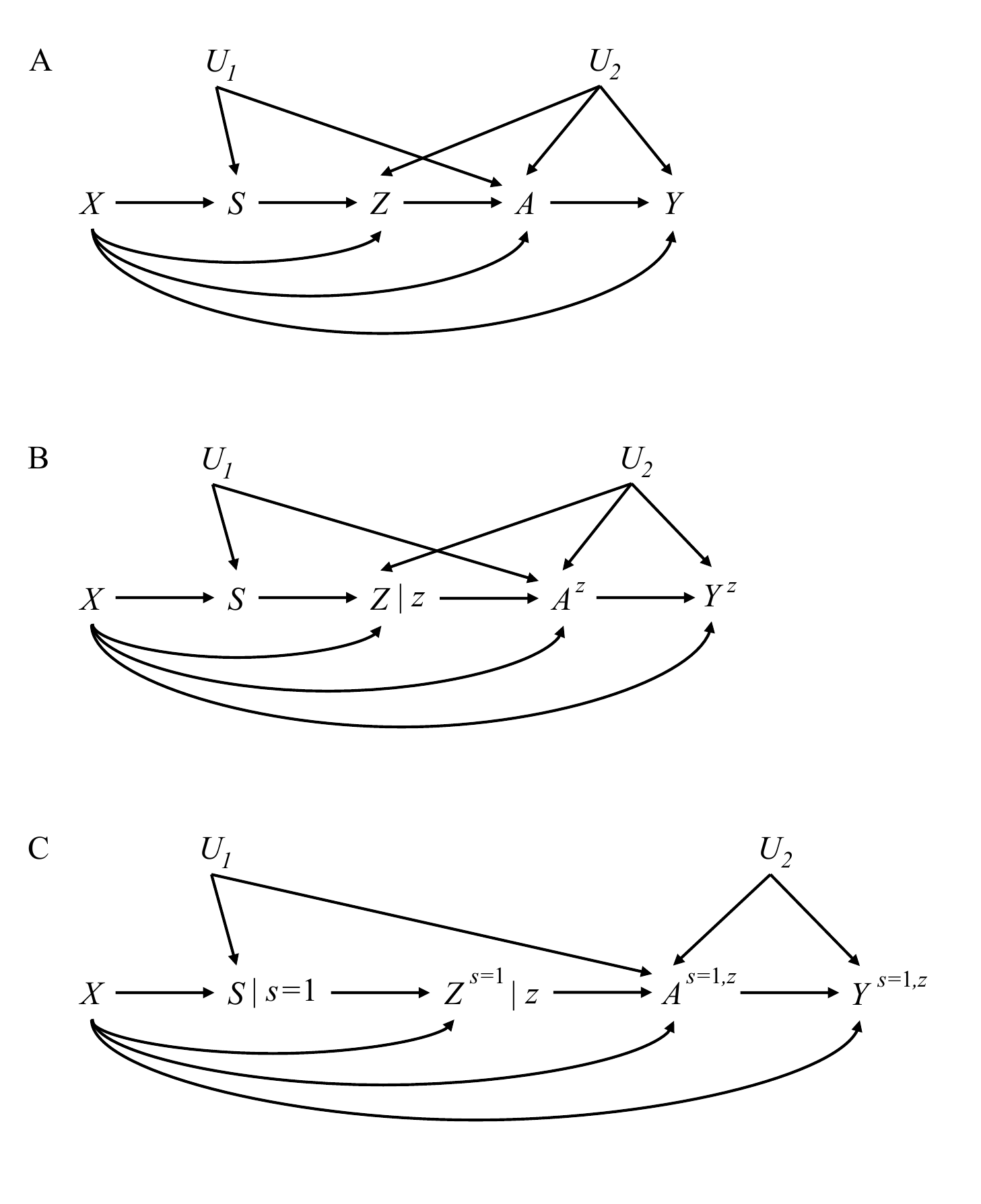}
    \caption*{}
    \label{fig:graphs_f3}
\end{figure}

\end{document}


\maketitle

\thispagestyle{empty}

\clearpage
\setcounter{tocdepth}{1}
\tableofcontents

\thispagestyle{empty}

\clearpage
\appendix 
\renewcommand{\thesection}{Web Appendix \arabic{section}}

\renewcommand{\thesubsection}{\arabic{section}.\arabic{subsection}}
\pagenumbering{arabic}

\renewcommand{\theequation}{\arabic{section}.\arabic{equation}}


\clearpage

\section{Identification in the target population}\label{appendix: direct_effect_of_participation}

\paragraph{Interventions to scale-up trial activities and recommend treatment:} We will show that the counterfactual means $\E[Y^{s=1,z}]$ in the target population, as well as average treatment effects comparing recommendations $z$ and $z^\prime$, $\E[Y^{s=1, z} - Y^{s=1, z^\prime}]$, are identifiable under the causal structure of Figure 1 in the main text. Starting from the DAG of Figure 1A in the main text, we construct the SWIG of Figure 1C under intervention to set $S$ to $s=1$ and $Z$ to $z$. The SWIG encodes two independence conditions: $Y^{s=1, z} \indep S|X$ and $Y^{s=1, z} \indep Z^{s=1} | (X, S)$. The second of these conditions, by consistency, implies that $Y^{s=1, z} \indep Z | (X, S=1).$

The above independence conditions, along with consistency and positivity, can be used to identify $\E[Y^{s=1,z}]$:
\begin{equation*}
    \begin{split}
    \E[Y^{s=1,z} ] &= \E \big[ \E[Y^{s=1,z} | X] \big] \\                 
        &= \E \big[ \E[ Y^{s=1,z} | X, S=1]  \big] \\
           &= \E \big[ \E[ Y^{s=1,z} | X, S=1, Z=z]  \big] \\
        &= \E \big[ \E[Y | X, S=1, Z=z]  \big].
    \end{split}
\end{equation*}
Here, the first equality follows from the law of total expectation; the second from $Y^{s=1, z} \indep S|X$; the third from $Y^{s=1, z} \indep Z | (X, S=1)$; the last from consistency; and all quantities are well-defined under the positivity conditions.

Furthermore, under positivity \cite{dahabreh2019relation}, 
\begin{equation*}
    \begin{split}
        \E \big[ \E[Y | X, S=1, Z = z] \big] &= \E\Bigg[ \E \left[ \dfrac{ I( S = 1 , Z = z) Y}{\Pr[S=1| X]\Pr[Z = z | X, S=1]} \Big | X  \right] \Bigg] \\
                &= \E\left[ \dfrac{ I(S=1, Z = z)Y}{\Pr[S=1| X]\Pr[Z = z | X, S=1]} \right], 
    \end{split}
\end{equation*}
where $I()$ is the indicator function that takes value 1 if $S = 1$ and $Z = z$, and 0 otherwise.

Using the above results, we can identify the average treatment effect comparing recommendation $z$ versus $z^\prime$, $\E[Y^{s=1,z} - Y^{z^\prime}]$, by taking differences: \begin{equation*}
    \begin{split}
    \E[Y^{s=1,z} - Y^{s=1, z^\prime}] &= \E \big[ \E[Y | X, S=1, Z=z] - \E \big[ \E[Y | X, S=1, Z=z^\prime] \\
    &= \E\left[ \dfrac{ I(S=1, Z = z)Y}{\Pr[S=1| X]\Pr[Z = z | X, S=1]} \right] - \E\left[ \dfrac{ I(S=1, Z = z^\prime)Y}{\Pr[S=1| X]\Pr[Z = z^\prime | X, S=1]} \right].
    \end{split}
\end{equation*}

\paragraph{Identification under the causal structure of Figure 2:} Under the causal structure of Figure 2, the conditions $Y^{s=1, z} \indep S|X$ and $Y^{s=1, z} \indep Z^{s=1} | (X, S)$ hold as under the causal structure of Figure 1. Thus, the above identification results also hold under the DAG of Figure 2.

\paragraph{Interventions to recommend treatment:} Under the causal structure of Figure 1, with perfect adherence, $Y^{s=1,z} = Y^{z}$. Thus the counterfactual means under the intervention to recommend treatment $z$ in the target population, $\E[Y^z]$, are equal to $\E[Y^{s=1,z}]$ and also equal to $\E \big[ \E[Y | X, S=1, Z=z] \big]$ and the weighting representation given above.

The situation is different, however, under the causal structure of Figure 2. Because of the $S \rightarrow A \rightarrow Y$ path on the DAG of Figure 1A in the main text, $Y^{s=1,z}$ is not in general equal to $Y^{z}$ \cite{dahabreh2019identification}. Thus, the counterfactual means under the intervention to recommend treatment $z$ in the target population, $\E[Y^z]$, is not in general equal to $\E \big[ \E[Y | X, S=1, Z=z] \big]$ or the weighting representation given above.

\paragraph{Simpler causal structures:} If the $S \rightarrow A$ edge or the $A \rightarrow Y$ edge was removed from the DAG of Figure 2A in the main text, then $Y^{z}=Y^{s=1,z}$ (this can be seen by modifying the DAG by removing the $S \rightarrow A$ edge or the $A \rightarrow Y$ edge and then constructing a new SWIG under joint intervention to set $S$ to $s=1$ and $Z$ to $z$ and using minimal labeling \cite{richardson2013single}). Under these modified causal structures, the identifiability result above holds if we substitute $Y^{z}$ for $Y^{s=1,z}$ (removal of either edge does not create additional dependencies).

\clearpage
\section{Identification in the non-randomized subset of the target population}\label{appendix:non_randomized_identification}

\paragraph{Interventions to scale-up trial activities and recommend treatment:} We will show that the counterfactual means $\E[Y^{s=1,z}| S = 0]$ in the non-randomized subset of the target population, as well as average treatment effects comparing recommendations $z$ and $z^\prime$, $\E[Y^{s=1, z} - Y^{s=1, z^\prime}| S = 0]$, are identifiable under the causal structure of Figure 1 in the main text. 

Starting from the DAG of Figure 1A in the main text, we construct the SWIG of Figure 1C under intervention to set $S$ to $s=1$ and $Z$ to $z$. As noted in \ref{appendix: direct_effect_of_participation}, we have two independence conditions: $Y^{s=1, z} \indep S|X$ and $Y^{s=1, z} \indep Z | (X, S=1)$. These conditions, along with consistency and positivity, can be used to identify $\E[Y^{s=1,z} | S = 0]$:
\begin{equation*}
    \begin{split}
    \E[Y^{s=1,z} | S =  0 ] &= \E \big[ \E[Y^{s=1,z} | X, S = 0] \big| S  = 0  \big] \\                 
        &= \E \big[ \E[ Y^{s=1,z} | X, S=1]  \big| S  = 0 \big] \\
           &= \E \big[ \E[ Y^{s=1,z} | X, S=1, Z=z] \big| S  = 0  \big] \\
        &= \E \big[ \E[Y | X, S=1, Z=z] \big| S  = 0  \big].
    \end{split}
\end{equation*}
Here, the first equality follows from the law of total expectation conditional on $S = 0$; the second from $Y^{s=1, z} \indep S|X$; the third from $Y^{s=1, z} \indep Z | (X, S=1)$; the last from consistency; and all quantities are well-defined under the positivity conditions.

Furthermore, under positivity \cite{dahabreh2019relation, dahabreh2020transportingStatMed},
\begin{equation*}
    \begin{split}
        \E \big[ \E[Y | X, S=1, Z = z] \big| S = 0  \big] &= \dfrac{1}{\Pr [ S = 0 ] } \E\Bigg[ I(S = 0)  \E \left[ \dfrac{ I( S = 1 , Z = z) Y}{\Pr[S=1| X]\Pr[Z = z | X, S=1]} \Big | X  \right] \Bigg] \\
        &= \dfrac{1}{\Pr [ S = 0 ] } \E\Bigg[   \E \left[ \dfrac{ I( S = 1 , Z = z) Y \Pr[S=0| X] }{\Pr[S=1| X]\Pr[Z = z | X, S=1]} \Big | X  \right] \Bigg] \\
        &= \dfrac{1}{\Pr [ S = 0 ] }  \E\left[ \dfrac{ I(S=1, Z = z)Y \Pr[S=0| X] }{\Pr[S=1| X]\Pr[Z = z | X, S=1]} \right].
    \end{split}
\end{equation*}

As in the previous sections of the Appendix, we can identify the average treatment effect comparing recommendations $z$ and $z^\prime$ in the non-randomized subset of the target population, $\E[Y^{s=1,z} - Y^{s=1, z^\prime} | S = 0]$, by taking differences between the identifying functionals of the corresponding counterfactual means.

\paragraph{Identification under the causal structure of Figure 2:} Under the causal structure of Figure 2A, the conditions $Y^{s=1, z} \indep S|X$ and $Y^{s=1, z} \indep Z^{s=1} | (X, S)$ hold as under the causal structure of Figure 1. Thus, the above identification results also causal structure under the causal structure of Figure 2.

\paragraph{Interventions to recommend treatment:} As discussed in the main text and \ref{appendix: direct_effect_of_participation}, under the causal structure of Figure 1, with perfect adherence, $Y^{s=1,z} = Y^{z}$. Thus the counterfactual means under the intervention to recommend treatment $z$ in the target population, $\E[Y^z | S = 0]$, are equal to $\E[Y^{s=1,z} | S = 0]$ and also equal to $\E \big[ \E[Y | X, S=1, Z=z] \big| S = 0 \big]$ and the weighting representation given above.

Also as discussed in the main text and \ref{appendix: direct_effect_of_participation}, the situation is different under the causal structure of Figure 2. Because of the $S \rightarrow A \rightarrow Y$ path on the DAG of Figure 2A in the main text, $Y^{s=1,z}$ is not in general equal to $Y^{z}$ \cite{dahabreh2019identification}. Consequently, the effect of recommendation in the non-randomized subset of the target population, $\E[Y^z | S = 0 ]$, is not in general equal to $\E \big[ \E[Y | X, S=1, Z=z] \big| S = 0 \big]$ or its weighting representation given above.

\paragraph{Simpler causal structures:} As discussed in \ref{appendix: direct_effect_of_participation} if the $S \rightarrow A$ edge or the $A \rightarrow Y$ edge was removed from the DAG of Figure 1A, then $Y^{z}=Y^{s=1,z}$
Under these modified causal structures, the identifiability result above holds if we substitute $Y^{z}$ for $Y^{s=1,z}$.

\paragraph{Non-nested trial designs:} The results for the identification of counterfactual means and average treatment effects in the non-randomized subset of the target population under nested trial designs, as presented in this section of the Appendix, can also be applied to non-nested trial designs \cite{dahabreh2021studydesigns}. In non-nested trial designs, the separate sampling of trial participants and non-participants induces a biased sampling model \cite{bickel1993efficient} (in the sense that the numbers of trial participants and non-participants under this design do not reflect the target population probability of trial participation) \cite{dahabreh2021studydesigns}. The results presented above are appropriate under the non-nested trial design provided the expectations and probabilities in the formulas are taken with respect to the biased sampling model \cite{dahabreh2021studydesigns, dahabreh2020transportingStatMed}.

\clearpage
\section{Identification in the population underlying the trial} \label{appendix:trial_population}

We will show that the counterfactual means for intervention to recommend treatment, $\E[Y^{z}|S=1]$ in the population underlying the trial (i.e., the randomized subset of the target population), as well as average treatment effects comparing recommendations $z$ and $z^\prime$, $\E[Y^{z} - Y^{z^\prime}|S=1]$, are identifiable under any of the DAGs presented in the main text of the paper. Consider the DAG of Appendix Figure 1, over both trial and non-trial participants, that combines the edges of the DAGs in Figures 1A and 2A from the main text, such that both a common cause of trial participation and a direct effect of participation on the received treatment are present. The DAG of Appendix 2A shows the corresponding DAG, conditional among trial participants (as indicated by the box around $S=1$). The $U_2 \rightarrow Z$ edge is removed because among trial participants, the randomized treatment recommendation is not confounded. Starting from the DAG of Appendix Figure 2A, we construct the SWIG of Appendix Figure 2B, conditional on $S=1$, under intervention to set $Z$ to $z$ and we obtain the independence condition $Y^z \independent Z | (X, S = 1)$.

This independence condition, along with the consistency and positivity conditions given in the main text, can be used to identify $\E[Y^z | S = 1]$:
\begin{equation*}
    \begin{split}
    \E[Y^z | S = 1] &= \E \big[ \E[Y^z | X, S = 1] \big | S = 1 \big] \\                 
        &= \E \big[ \E[ Y^z | X, S =1, Z = z ] \big | S = 1 \big] \\
        &= \E \big[ \E[Y | X, S= 1, Z = z] \big | S = 1 \big].
    \end{split}
\end{equation*}
Here, the first equality follows from the law of total expectation; the second from $Y^z \independent Z | ( X, S = 1) $; the last from consistency; and all quantities are well-defined under the positivity condition for treatment in the trial.

Furthermore, under positivity \cite{hernan2006estimating, dahabreh2019relation}, 
\begin{equation*}
    \begin{split}
        \E \big[ \E[Y | X, S = 1, Z = z] \big | S = 1 \big] &= \E\Bigg[\E\left[ \dfrac{ I(S=1, Z = z)Y}{\Pr[Z = z | X, S = 1]} \Big| X, S = 1  \right] \Bigg| S = 1\Bigg]  \\
            &= \E\left[ \dfrac{ I(S=1, Z = z)Y}{\Pr[Z = z | X, S = 1]} \Big| S = 1  \right].
    \end{split}
\end{equation*}

As in the previous section of the Appendix, we can identify the average treatment effect comparing recommendations $z$ and $z^\prime$ in the target population underlying the trial, $\E[Y^z - Y^{z^\prime} | S = 1]$, by taking differences between the identifying functionals of the corresponding counterfactual means.

\paragraph{Simpler causal structures:} If we simplify the DAG of Appendix Figure 1, by either removing the $U1 \rightarrow S$ and $U1 \rightarrow A$ (resulting in a DAG similar to that of Figure 2A in the main text) or by removing the $S \rightarrow A$ edge (resulting in the DAG of Figure 3A in the main text), the identifiability results above still hold 
(removal of these edges does not create additional dependencies).

\clearpage
\section{Figures}
\renewcommand{\figurename}{Appendix Figure}\label{appendix:figures}

\begin{figure}[htp]
 \caption{DAG for for direct effect of participation on adherence and common causes of trial participation and treatment.}
    \vspace{0.4in}
    \centering
       \includegraphics[width=11cm]{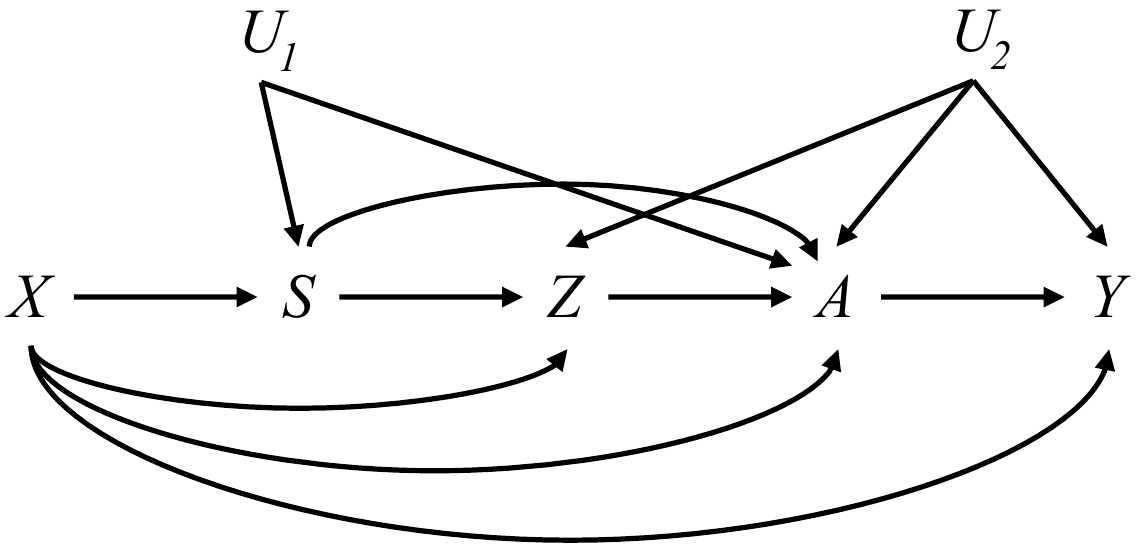}

    \caption*{}
    \label{fig:graphs_f2}
\end{figure}

\begin{figure}[htp]
 \caption{DAG and SWIG for the population underlying trial.}
    \vspace{0.4in}
    \centering
    \includegraphics[width=11cm]{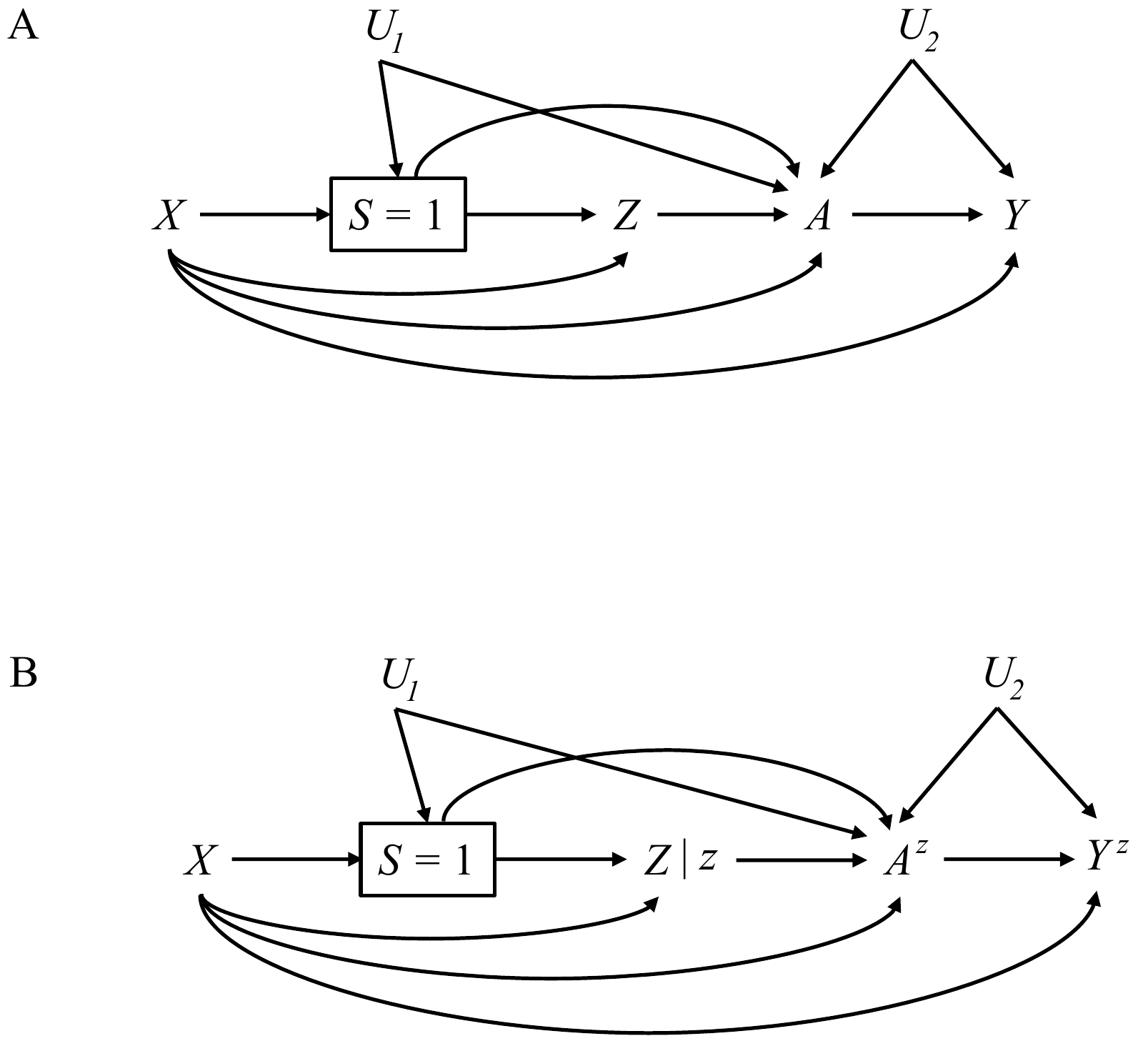}

    \caption*{}
    \label{fig:graphs_f1}
\end{figure}

\clearpage
\bibliographystyle{unsrt}
\bibliography{bibliography}{}